# Genetic Analysis of Transformed Phenotypes

Nicolo Fusi[1,*], Christoph Lippert[1], Neil D. Lawrence[2], Oliver Stegle[3,*]

[1] eScience group, Microsoft Research, Los Angeles, USA
[2] Department of Computer Science, University of Sheffield, Sheffield, UK
[3] European Molecular Biology Laboratory, European Bioinformatics Institute, Hinxton, Cambridge, UK
[*] To whom correspondence should be addressed: fusi@microsoft.com, stegle@ebi.ac.uk

Linear mixed models (LMMs) are a powerful and established tool for studying genotype-phenotype relationships. A limiting assumption of LMMs is that the residuals are Gaussian distributed, a requirement that rarely holds in practice. Violations of this assumption can lead to false conclusions and losses in power, and hence it is common practice to pre-process the phenotypic values to make them Gaussian, for instance by applying logarithmic or other non-linear transformations. Unfortunately, different phenotypes require different specific transformations, and choosing a "good" transformation is in general challenging and subjective. Here, we present an extension of the LMM that estimates an optimal transformation from the observed data. In extensive simulations and applications to real data from human, mouse and yeast we show that using such optimal transformations leads to increased power in genome-wide association studies and higher accuracy in heritability estimates and phenotype predictions.

## Introduction

Linear mixed models (LMMs) are widely used in genetic studies on humans and a variety of model organism. This model class is attractive because in addition to the effects of single genetic variants, they can account for polygenic effects and confounding due to population structure or family relatedness. Important applications of linear mixed models in genetics include genome-wide association studies[1,2], narrow-sense heritability estimation[3,4] and phenotype prediction[5–8].

One of the core assumptions of LMMs is that the noise distribution is Gaussian, and deviations from Gaussianity can result in model misspecification[9]. It is standard practice to apply a transformation to the phenotype in cases in which this assumption may be violated. For instance, if the scale of the phenotype spans several orders of magnitude, it is common to log-transform it before performing genetic analyses. Log transformations are also a popular choice when the phenotypic measurement is defined as ratio between a foreground and a background signal, such as in gene expression measurements from microarrays or when analyzing composite phenotypes (e.g. the ratio between total cholesterol and high density lipoprotein). Nonetheless, the set of transformations commonly used in genetic studies is extremely rich[10–13] and no single transformation can be considered a universal solution. For instance, a recent study of 58 different mouse traits[14] proposed the selection of a separate transformation function for each trait. In this context, manually choosing transformation functions has two drawbacks. First, there is no concrete quantitative way of choosing a transformation over another. This is because the objective is not to obtain Gaussian distributed phenotypes, but rather Gaussian distributed noise, or equivalently, Gaussian distributed residuals after fitting an unknown genetic model. The second drawback is that the number of possible transformation functions that can be manually explored is limited. Exhaustively testing different parameterizations of several transformations functions is time consuming and can result in a multiple-hypothesis-testing problem, since the same analysis is repeated multiple times under different transformations.

Here we propose the warped linear mixed model (WarpedLMM), a principled generalization of the standard LMM in which the transformation function is learned directly from the data. We show how the likelihood principle allows to objectively assess alternative transformations in the light of the observed genotype and phenotype data. WarpedLMM can seamlessly be used in place of traditional LMMs Moreover, the transformations inferred by WarpedLMM are parametric and invertible, thus permitting to predict phenotypic values on the original scale. This is not possible, for instance, when considering non-parametric transformations with rank statistics.

We perform an extensive investigation of the performance of WarpedLMM in key applications in genetics across 50,000 simulated datasets, as well as on real data from human, mouse and yeast. We compare WarpedLMM to established techniques such as Box-Cox transformations or rank transformations in combination with a LMM, demonstrating that WarpedLMM more accurately recovers the true underlying transformations. Overall, we show that using a WarpedLMM can be used in place of a standard LMM for a wide range of genetic analyses, resulting in an increase of power in GWAS, a reduction of bias in narrow-sense heritability estimates and an increase in phenotype prediction accuracy.

## Results

In our model (Supplementary Figure 1), the observed phenotype is determined by applying a non-linear transformation function **f** to the latent phenotype. Thus, in order to recover the true genetic model, an estimate of the inverse transformation $\mathbf{f}^{-1}$ is needed. WarpedLMM builds on the assumption that this transformation can be approximated by an invertible parametric "warping" function (see Methods). The behavior of this warping function is determined by a small number of parameters that are treated as additional model parameters in a LMM. The most probable transformation can then be determined by maximizing the sum of the log-likelihood and a regularization term that penalizes the complexity of the fitted invertible function.

## Simulations

First, we considered the problem of narrow-sense heritability estimation on simulated data, where ground truth information is available.

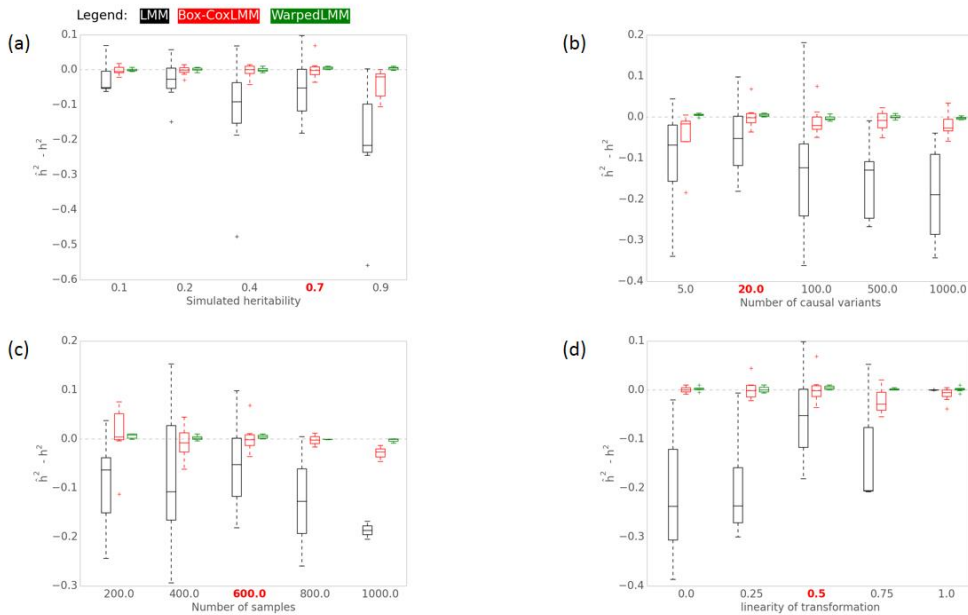

*Figure 1*: Simulation experiment comparing different LMM approaches for estimating the genetic proportion of phenotype variability (narrow-sense heritability, $h^2$). (**a**) changing the simulated heritability (**b**) considering different numbers of causal variants (**c**) increasing the sample size and (**d**) decreasing the non-linearity of the true simulated transformation (at 0 the function is completely non-linear, while at 1.0 is completely linear and no transformation is needed. See Methods for details). For each parameter, the remaining simulation settings remained constant with the default parameters being highlighted in red. Heritability estimates were obtained using WarpedLMM, a LMM, and a LMM on Box-Cox preprocessed phenotypes.

We simulated phenotypic effects based on human genotype data from the HapMap project[15]. We performed multiple simulations changing the proportions of variance explained by the genotype, the number of causal variants and the observed sample size. In each simulation experiment, we generated the observed phenotype by applying a transformation function to the simulated phenotype. In an effort to keep our simulations as realistic as possible, we considered the same transformations identified on real data from mouse by Valdar et al.[14]. Additionally, the transformation function was controlled by a parameter that controlled the degree of non-linearity, interpolating between a linear function (no transformation) and a completely nonlinear function (full transformation). Based on transformed phenotype, we then compared the ability of the WarpedLMM and the LMM to recover the true simulated heritability. We also applied a LMM using a phenotype transformed using a Box-Cox transformation[16] (Box-CoxLMM), which is commonly used in practice[16–21] as an alternative to manually chosen transformations.

When comparing the estimated heritability to the true simulated one, WarpedLMM consistently reported highly accurate heritability estimates, while the LMM consistently underestimated heritability.

WarpedLMM correctly estimated the true simulated heritability irrespective of the heritability level (Figure 1a), number of causal variants (Figure 1b), number of samples (Figure 1c) or linearity of the transformation (Figure 1d). Strikingly, we also observed that increasing the number of samples considered in the study (Figure 1c) did not reduce the estimation error of the LMM. Likewise, we found the accuracy of the heritability estimates from a LMM to be negatively affected by the true underlying heritability (Figure 1a) and the number of causal variants (Figure 1b). Not surprisingly, the degree of non-linearity of the transformation had the largest effect on the model accuracy (Figure 1d), where even subtle non-linearity of the transformation functions had a profound effect on the model estimates. It should be noted that, even when the transformation is completely linear (rightmost point in Figure 1d) and thus no transformation is needed, WarpedLMM achieved approximately the same estimation error as a standard LMM, demonstrating that the method is robust and can be used even in settings where no transformation is needed. Additional results for different transformations and extensive comparisons to other methods are show in Supplementary Figures 2 and 3.

## Mouse data from Valdar et al.

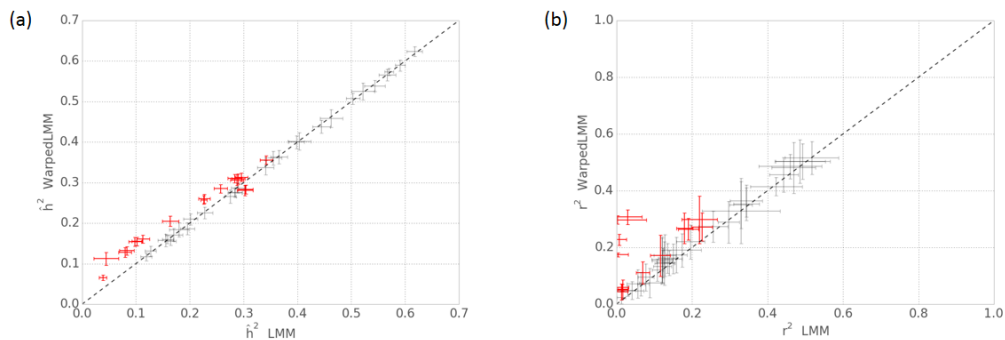

*Figure 2* Comparative analysis of WarpedLMM and a LMM on the mouse dataset. Panel (**a**) shows heritability estimates using a LMM on the untransformed phenotype versus the heritability estimates obtained by WarpedLMM. Empirical error bars were obtained from 10 bootstrap replicates, using 90 % of the data in each replicate. Significant differences are colored in red (paired t-test, α = 0.05). Panel (**b**) shows out-of-sample prediction accuracy assessed by the squared correlation coefficient $r^2$, considering either a LMM on the untransformed data and a WarpedLMM. Prediction accuracies were assessed from 10 random train-test splits. Phenotypes with significant deviations in prediction accuracy of the LMM and the WarpedLMM are highlighted in red (paired t-test, p-value ≤ 0.05).

Next, we revisited data from a heritability study in a structured mouse population[14]. This study was one of the motivations of this work on automating phenotype transformation, because it showed that accurate association results depend on carefully defining a specific transformation for each of the 47 phenotypes under consideration. While this process was guided by an initial Box-Cox fit, the authors performed further manual tuning of the resulting function for each phenotype independently. Here, we compared a LMM on untransformed phenotypes to estimates derived using WarpedLMM. Covariates such as age, gender, body weight, litter number and cage density were included as fixed effects in both models. We found that the two models yielded significantly different heritability estimates (Figure 3b, p-value ≤ 0.05 from a paired t-test) for 18 of the 47 phenotypes. For most of these (17 out of 18) WarpedLMM yielded a higher estimate of narrow sense heritability than a standard LMM.

Unlike the simulated experiments described in the previous section, we lack an accurate gold standard to validate the heritability estimates on real data. To this end, we validated our findings by comparing both models in an out-of-sample prediction task. We performed 10-fold cross validation, where each models is repeatedly trained on 90% of the data to predict the phenotype from genotype on the remaining 10% of the samples. WarpedLMM consistently yielded more accurate out-of-sample predictions than a standard LMM (Figure 3d), even for phenotypes where the estimated heritability was lower (Supplementary Figure 5b). This suggests that appropriate phenotype transformations help avoiding under or overfitting in applications of mixed models, confirming our results on simulated data and supporting that the heritability estimates of WarpedLMM are also more accurate on real data.

Finally, when comparing the transformations identified by WarpedLMM to those manually derived by Valdar et al.[14], we found that the functions estimated by WarpedLMM were consistently in the same class (linear, logarithmic, etc.) as those reported in the original study, however with slight differences in parameterization (Supplementary Figure 6).

Supplementary figures 4a-b and 5a provide equivalent results for a similar study in yeast, demonstrating that these findings hold also for other systems.

## WarpedLMM for GWAS

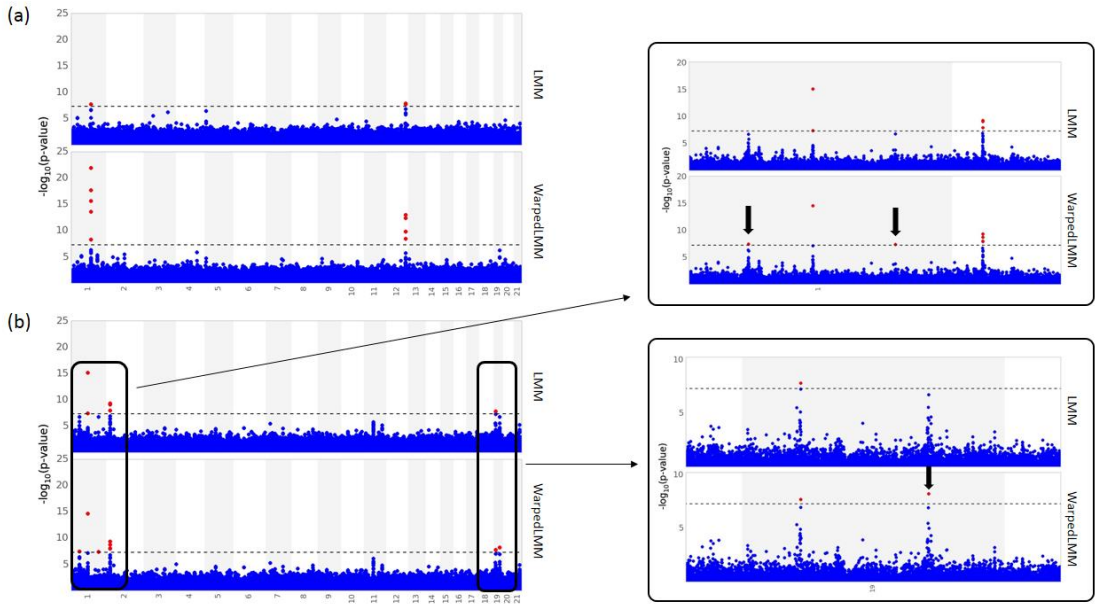

*Figure 3* Manhattan plots of a GWAS of (**a**) C-reactive protein (CRP) and (**b**) low-density lipoprotein (LDL) using a LMM applied to untransformed phenotypic values and WarpedLMM. Red circles represent significant associations at a significance level of $5 \times 10^{-8}$ (marked on the plots with a dashed line). The two rightmost panes show an enlarged view of interesting regions in chromosomes 1 and 19, with black arrows highlighting loci that were identified only when using WarpedLMM.

In addition to heritability estimation and prediction, WarpedLMM can also be used to perform genome-wide association studies. To test this, we revisited genotype and phenotype data from the Northern Finnish birth cohort[22] and analyzed four related metabolic traits: high density lipoprotein (HDL), low density lipoprotein (LDL), triglycerides (TRI) and C-reactive protein (CRP). This selection of four phenotypes is particularly interesting because, although the phenotypes are closely related in mechanism, in the initial publication[22] it has been proposed to log transform some of the phenotypes (TRI, CRP) while leaving the remaining phenotypes (HDL, LDL) on the original scale.

We carried out a univariate GWAS using three different methods: WarpedLMM, a LMM applied to untransformed phenotypes[1] and a LMM on phenotypes transformed as reported in the original paper[22]. Association results from all methods appropriately controlled for type 1 error rate (genomic control for all methods was 1.00 ± 0.01). Overall, using WarpedLMM resulted in an increase in power to detect associations (Supplementary Table 1). For example, WarpedLMM identified a total of 6 distinct loci that were significantly associated (p-value ≤ $5\times10^{-8}$) to LDL cholesterol levels (Figure 3b), while all the other methods only identified 3. Notably, of these three new loci, two have been identified in previous studies. In particular rs4844614 has been significantly associated with LDL in an analysis of the same data using linear regression[22] and rs4844614 has been identified in a large meta-analysis[23].

Similarly, WarpedLMM identified 3 QTLs for HDL cholesterol, while all the other methods missed one of these QTLs. Even in cases in which no new locus was identified, such as in the analysis of CRP, WarpedLMM was more sensitive in picking up the genetic signals when compared with a standard LMM (Figure 3a).

Furthermore, we found that separate application of WarpedLMM to each of the 4 phenotypes increased pairwise correlations structure between phenotypes, which is important for multivariate analyses[24,25] (Supplementary Figure 8). Indeed, semi-parametric transformation approaches have previously been applied for multivariate analyses[26] on this dataset. In particular, these approaches consisted in rank-standardizing individual phenotypes prior to regressing out covariates, followed by an additional rank-standardization step[26]. The assumption behind this approach is that the genotype explains only a small proportion of the variance and that the covariates contain most of the confounding signal for recovering the correct transformation. While this may not be true in general, we found this assumption to be realistic for this specific dataset, as evidenced by a comparison of the transformations recovered by WarpedLMM and by the semi-parametric approach of Zhou and Stephens[26]. Indeed, we observed striking correlations between both the functions recovered (Supplementary Figure 7) and the p-values obtained by the two methods when used in a univariate GWAS on each trait (ρ = 0.99 ± 0.01, Figure 4).

Finally, we validated the full genetic model implied by WarpedLMM using out-of-sample phenotype prediction. Since the transformations functions found by WarpedLMM can be inverted, it is possible to assess prediction accuracy on the natural scale, unlike when using rank-based preprocessing methods[26]. We observed a consistent improvement in out-of-sample prediction when employing WarpedLMM compared to a standard LMM, suggesting that it accurately models the phenotype data (Supplementary Table 1). Overall, these experiments support that WarpedLMM can be used as a robust preprocessing procedure for GWAS.

## Discussion

Although preprocessing methods are widely used in practice to invert an unknown phenotype transformation[10–13,17,19–21,26–28], so far there has been no principled approach to assess and fit different transformations while accounting for genetic information and covariates.

Here, we have shown how the classical LMM can be extended to estimate phenotype transformations directly from the data. Our experiments show that WarpedLMM is able to significantly improve the accuracy and power of important genetic analyses, including heritability estimation, prediction and GWAS. Although an important application of WarpedLMM is the generation of transformed phenotypes for downstream analysis, we emphasize that the model is much more than an ad hoc pre-processing procedure. The objective function of the model can be derived from first principles, resulting in an extension of the mixed model to balance the data likelihood and the complexity of the fitted transformation (Methods). As a result, our approach can be directly applied to tasks commonly tackled using linear mixed models, such as GWAS, heritability estimation and phenotype prediction.

When applying WarpedLMM to studies in mouse and yeast, we found an overall increase of the proportion of variance that could be attributed to genetic factors. Although in a minority of traits the heritability estimates decreased, we note that the model consistently improved out-of-sample prediction. This shows that inappropriate phenotype transformations can lead to overoptimistic heritability estimates and overfitting, a fact that has previously been noticed by others[29]. Remarkably, although WarpedLMM has a larger number of parameters than a standard mixed model, it did not overfit even for sample sizes (Figure 1a) that are much smaller than the ones used in typical studies.

Although we have focused on some of the most established tasks in genetic analysis, WarpedLMM can easily be used in more specialized analyses. For example, it is possible to use the model in combination with multi locus mixed models[30] or mixed models that jointly consider multiple phenotypes[24,25]. WarpedLMM finds the transformation function while jointly taking into account all the available covariates and the genotype data. This joint approach helps to ensure that the model residuals are Gaussian distributed, rather than the phenotype itself. The importance of this principle has been recognized in previous work[26], in which the authors employed a three-step procedure which consisted of rank transforming the phenotype, regressing out the covariates and rank transforming the residuals again. This approach assumes that the genotype explains only a small portion of the variance and hence Gaussianizing phenotype data on the null model is valid. While this approach is reasonable in some analyses, deviations from this assumption remain a concern[28] and highlight the need for principled approaches such as WarpedLMM that put this principles on solid statistical grounds.

Finally, we note that there may be scenarios where also WarpedLMM does not achieve optimal results. Similar to other existing methods, the model learns a transformation but assumes that that the noise level in the transformed phenotype space is constant. This assumption may be violated in some cases such as when dealing with count data or binary phenotypes. In such instances, it will remain appropriate to use generalized linear mixed models with non-Gaussian likelihoods that incorporate stronger assumptions about the nature of the data. Nonetheless, the number of phenotypes being measured is constantly increasing and only a small fraction will obey well defined properties of either being binary or Poisson distributed. In these instances there are clear advantages of the WarpedLMM model: it allows robust analyses of a broad spectrum of phenotypes without the need to develop specialized methods or carry out manual inspection of the transformations

## Methods

We model the observed non-normal distributed phenotype $y_n$ of each individual $n$ with an unobserved normal distributed phenotype $z_n$ that results from transforming $y_n$ using the monotonic function $f$ with some parameters $\psi$.

$$z_n = f(y_n; \psi)$$

On the normal distributed scale, the representation $z_n$ of the phenotype is given by the following linear mixed model

$$z_n = \mathbf{x_n}\boldsymbol{\beta} + \mathbf{g_n^*}\boldsymbol{\alpha} + \epsilon_n \tag{1}$$

Where $x_n$ holds the covariates for individual $n$, $\boldsymbol{\beta}$ are fixed effects, $\boldsymbol{g_n^*}$ contains the genotype of the individual at $S^*$ genetic loci, $\boldsymbol{\alpha}$ are normal distributed random genetic effects and $\epsilon_n$ is independent normal distributed noise.

Given this linear mixed model, the likelihood for N-by-1 vector $\mathbf{z} = f(\mathbf{y}; \boldsymbol{\psi})$ of transformed phenotypes for a sample of N individuals is

$$\mathbf{z} \sim N(\mathbf{X}\boldsymbol{\beta}, \sigma_g^2 \mathbf{K} + \sigma_e^2 \mathbf{I}), \tag{2}$$

Where $\mathbf{K}$ is the relationship matrix at the causal loci, $\sigma_g^2$ is the total amount of genetic variance and $\sigma_e^2$ is the error noise variance.

In practice we use a genomic relatedness matrix[31] computed from all S genotyped common SNPs, pre-processed to have zero mean and unit variance and stored in the $N \times S$ matrix $\mathbf{G}$

$$\mathbf{K} = \frac{1}{S} \mathbf{G}\mathbf{G}^\mathsf{T}$$

### Choosing a monotonic warping function

Instead of specifying a fixed transformation, we find the optimal transformation $\tilde{f}$ for a given dataset by maximizing the likelihood (3) of the transformed phenotype over a flexible class of monotonic functions parameterized by $\psi$.

Following Snelson et al[32]., for the phenotype $y_n$ of each sample, the transformation is chosen as

$$f(y_n; \psi) = d \cdot y_n + \sum_{i=0}^{I} a_i + \tanh(b_i \cdot (y_n + c_i)) \qquad a_i \geq 0, b_i \geq 0, d \geq 0, \qquad \forall i$$

where $\psi = (d, a_1, b_1, c_1, \ldots, a_I, b_I, c_I)$.

In this equation, $f$ is a sum over $I$ non-linear step functions, where each $a_i$ controls the step size, $b_i$ controls the steepness and $c_i$ controls the location. Additionally, the parameter $d$ is a coefficient for the linear part (in $y_n$) of the function.

The only parameter requiring to be set manually is the number $I$ of step functions. We followed the recommendation in Snelson et al. and used $I = 3$ step functions for all of our experiments.

## Parameter estimation

The model parameters are estimated by maximizing a penalized form of the linear mixed model likelihood. By taking the logarithm of (3), the negative log likelihood $L$ for the hidden normal distributed phenotype $\mathbf{z}$ is obtained as

$$L = -\log P(\mathbf{z} \mid \mathbf{X}, \mathbf{G}) = \frac{1}{2}\log \det \mathbf{C}_N + \frac{1}{2}(\mathbf{z} - \mathbf{X}\boldsymbol{\beta})^\top \mathbf{C}_N^{-1}(\mathbf{z} - \mathbf{X}\boldsymbol{\beta}) + \frac{N}{2}\log 2\pi.$$

The previous equation is not accounting for the fact that $\mathbf{z}$ is really a transformation of the observed phenotype $\mathbf{y}$. This transformation can be taken into account by including the corresponding Jacobian term, yielding the negative log likelihood for $\mathbf{y}$ as

$$L = \frac{1}{2}\log \det \mathbf{C}_N + \frac{1}{2}(f(\mathbf{y}; \boldsymbol{\psi}) - \mathbf{X}\boldsymbol{\beta})^\top \mathbf{C}_N^{-1}(f(\mathbf{y}; \boldsymbol{\psi}) - \mathbf{X}\boldsymbol{\beta}) - \sum_{n=1}^{N}\log \frac{\partial f(\mathbf{y}; \boldsymbol{\psi})}{\partial \mathbf{y}} + \frac{N}{2}\log 2\pi. \quad (3)$$

It is then possible to fit the model by minimizing (3) with respect to the parameters of the model and the transformation.

## Incorporating strong genetic effects

While the realized relationship matrix $\mathbf{K}$ can accurately capture the relatedness between individuals in the presence of many causal variants with small effect sizes, it doesn't necessarily do so when the genetic signal is mostly due to a small number of causal variants. For this reason, several approaches[30,33,34] have been proposed to select strong genetic effects for inclusion in the model. Here, we perform a forward selection procedure[33,34] by iteratively adding a new variance component representing the strongest effect to the random effects term.

At iteration $t$ is thus defined as

$$\mathbf{z} \sim N\left(\mathbf{X}\boldsymbol{\beta}, \sigma_k^2 \mathbf{K} + \sum_{i=1}^{t} \sigma_i^2 \mathbf{G}_i \mathbf{G}_i^\top + \sigma_e^2 I\right),$$

where the parameters $\boldsymbol{\psi}, \boldsymbol{\beta}, \sigma_g^2, \sigma_i^2, \sigma_e^2$ are re-estimated at each iteration.

In each iteration $t$, the SNP with the strongest individual effect is determined by fixed effects testing[2] of all genetic markers against the current transformed phenotype $\mathbf{z}_t$ using the current set of variance components as the relatedness matrix. A marker is selected if its q-value[35] is smaller than a threshold, which we set to 0.05 for all our experiments. The algorithm converges when no marker achieves genome-wide significance at the FDR level specified.

The genetic effects incorporated in the model at the end of this procedure can in general be beneficial for certain tasks such as phenotype prediction. Here we only use them to better reconstruct the transformation function, and we do not take them into account while doing prediction or heritability estimation. Finally, it is important to notice that alternatives to the forward selection technique described here can be used to select the genetic variants to be included in the model.

### Phenotype prediction

Under this model we can predict the unobserved phenotype of a new individual indexed by * given the genotype alone. Assuming a fully observed sample of N individuals, we can use the parameter estimates under model (2) to compute the best linear unbiased predictor (BLUP) $\hat{z}_*$ of the new individual's phenotype on the normal distributed scale

$$\hat{z}_* = \mathbf{x}_*\boldsymbol{\beta} + \hat{\sigma}_g^2 \mathbf{k}_* \left(\hat{\sigma}_g^2 \mathbf{K} + \hat{\sigma}_e^2 \mathbf{I}\right)^{-1}(\mathbf{z} - \mathbf{X}\boldsymbol{\beta}),$$

where $\mathbf{x}_*$ is a vector of covariates for the new individual, $\mathbf{k}_*$ is a 1-by-N vector that contains the genomic relatedness between the new individual and all the individuals in the original sample.

In order to get an estimate of the phenotype on the original scale, we apply the reverse transformation $f^{-1}$ to the best linear unbiased predictor

$$\hat{y}_* = f^{-1}(\hat{z}_*; \hat{\psi})$$

The reverse transformation $f^{-1}$ is obtained by numerically inverting $f$ using Newton-Raphson updates as done by Snelson et al.

### Estimating heritability

We obtain an estimate of the narrow-sense heritability $h^2$ in the normal distributed scale by computing a chip heritability $\hat{h}^2$ from common genotyped markers in the linear mixed model (2).

$$\hat{h}^2 = \frac{\hat{\sigma}_g^2}{\hat{\sigma}_e^2 + \hat{\sigma}_g^2},$$

where $\hat{\sigma}_g^2$ and $\hat{\sigma}_e^2$ are restricted maximum likelihood (REML) estimates of $\sigma_g^2$ and $\sigma_e^2$.

### Simulation study

The simulated data is generated taking genotypes from hapmap3[15] chromosome 22. In each simulation, we sample an $h^2$ from {0.1,0.20,0.40,0.70,0.9}, the number of causal variants from {5,20,100,500,1000}, the number of samples from {200,400,600,800,1000}, the variance explained by covariates from {0.0,0.25,0.5,0.70,0.9}. We can then recover the noise level conditioned on $h^2$, and the covariates variance.

Finally, we pick a transformation $f(y)$ from the set of transformations used in Valdar et al.[14] (for the experiments in the main paper we used $\exp(y)$, other transformations are available in the supplementary material). We then transform the phenotype as $z = t \cdot y + (1 - t)f(y)$, where $t$ is a parameter that determines the intensity of the transformation and is sampled from {0.0 , 0.25, 0.5, 0.75, 1.0}. We repeated this simulation procedure 50,000 times in order to have a sufficiently large sample size to investigate all the regimes described above.

### Mouse data

We used mouse data from Valdar et al.[14]. This dataset contains between 1700 and 1940 samples (depending on phenotype missingness), 10,132 markers and 47 phenotypes.

### Human data

We used the data from Sabatti et al.[22] and applied the same filtering criteria described in Zhou et al.[26]. This resulted in 5,255 individuals and 328,517 SNPs.

# Supplementary Material

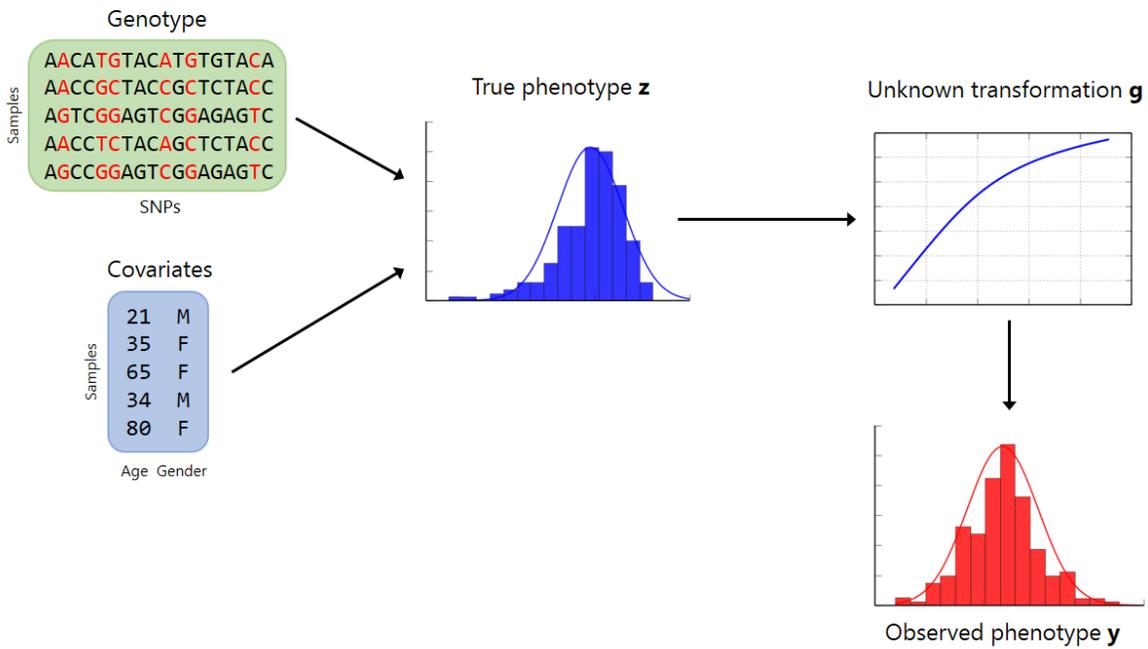

*Supplementary Figure 1* The genetic model of interest determines the latent phenotype profiles **z** (blue histogram), the measured phenotype data **y** (red histogram) are then derived from **z** via an unknown transformation **f**.

We repeated the simulation experiments described in the main paper using different phenotype transformations and comparing several different models. To keep our simulations realistic, we only used transformations found in real data (Valdar et al., 2006)

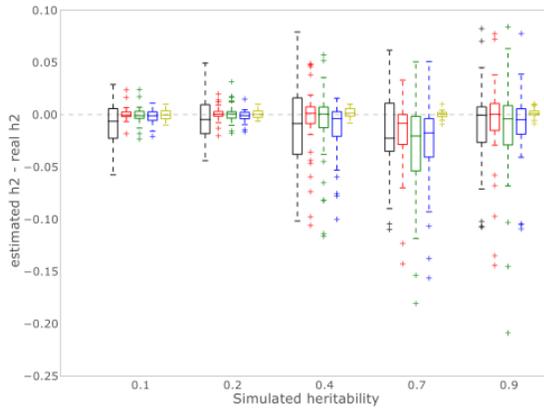

(a) Varying the simulated heritability

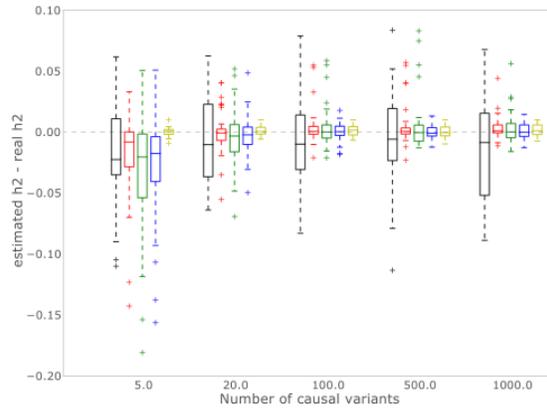

(b) Varying the number of causal variants

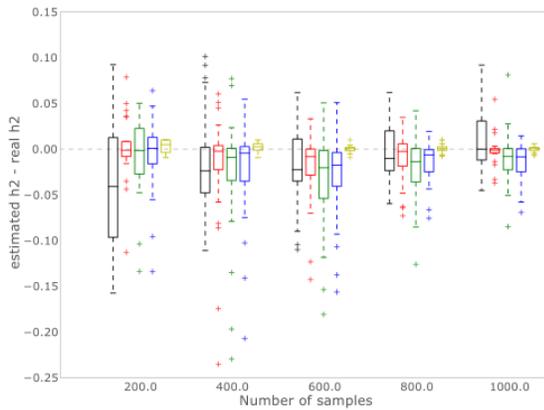

(c) Varying the number of samples

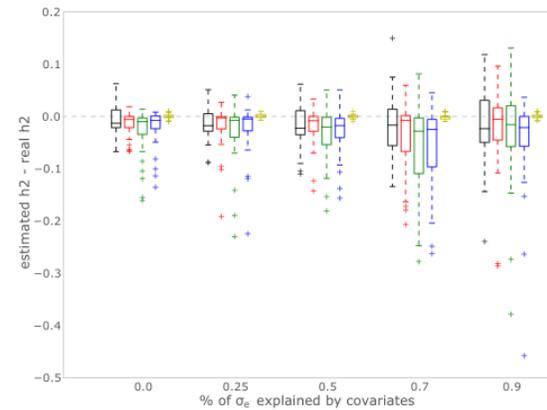

(d) Varying the intensity of the transformation

*Supplementary Figure 2* Comparison of alternative linear mixed-model approaches for estimating the genetic contribution to phenotype variability (narrow sense heritability, $h^2$). As done in the main paper, we evaluate the difference between the estimated and the true genetic variance across 50'000 simulated experiments. In this particular experiment we considered a different transformation ($z = \sqrt{y}$) and included comparisons to a rank-based transformation and a simpler version of the WarpedLMM model which incorporates genetic information with a full rank kernel only (realized relationship matrix). Legend: **LMM**, **Box-Cox**, **WarpedLMM**, **WarpedLMM with full RRM only**, **Rank transformation**

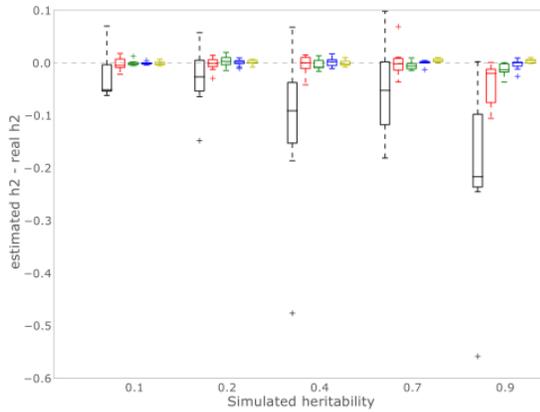
(a) Varying the simulated heritability

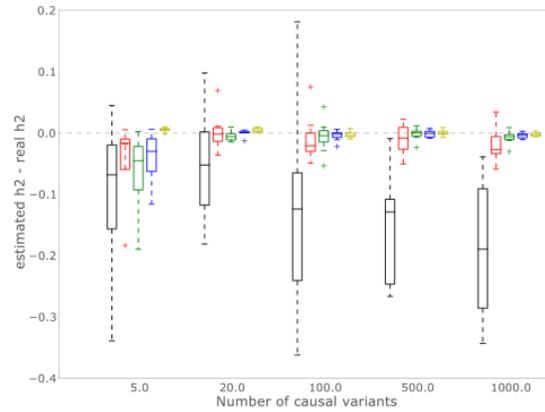
(b) Varying the number of causal variants

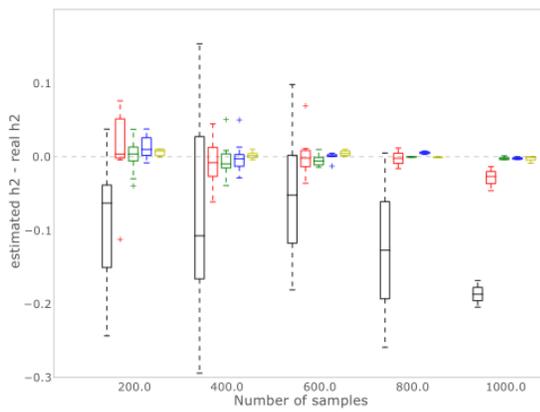
(c) Varying the number of samples

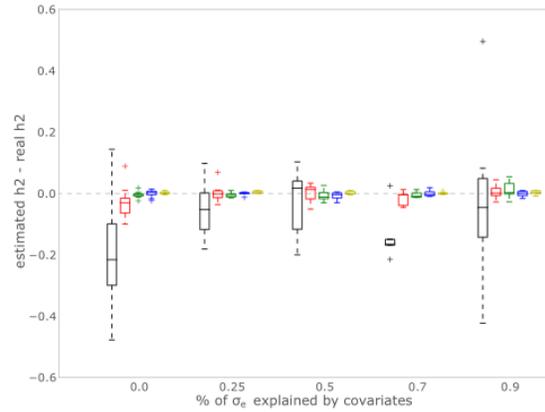
(d) Varying the intensity of the transformation

*Supplementary Figure 3* Comparison of alternative linear mixed-model approaches for estimating the genetic contribution to phenotype variability (narrow sense heritability, $h^2$). As done in the main paper, we evaluate the difference between the estimated and the true genetic variance across 50'000 simulated experiments. Here, we considered the transformation $z = exp(y)$ and included comparisons to a rank-based transformation and a simpler version of the WarpedLMM model which incorporates genetic information with a full rank kernel only (realized relationship matrix). Legend: **LMM**, **Box-Cox**, **WarpedLMM**, **WarpedLMM with full RRM only**, **Rank transformation**

## 2) Analysis of yeast data from Bloom et al.

Next, we considered a study on a F2 yeast cross (Bloom, Ehrenreich, Loo, Lite, & Kruglyak, 2013), to understand the implication of phenotype transformation in a well-powered study with highly heritable traits. Figure 3a shows narrow-sense heritability estimates using a standard linear mixed model versus heritability estimates using transformations fitted by WarpedLMM. These methods results in significantly deviating heritability estimates (paired t-test, α = 0.05) for 17 phenotypes (38%), most of which with increased heritability by WarpedLMM compared to the standard approach (11 of 17, 65%). This suggest that even phenotypes obtained in controlled settings tend to be transformed, leading to both overestimation and underestimation of the narrow-sense heritability. To validate the genetic models derived using WarpedLMM, we performed out-of-sample phenotype prediction using both a WarpedLMM and a standard LMM (Supplementary figure 3b). Reassuringly, the WarpedLMM model consistently yielded improved prediction accuracy, irrespective of whether the heritability estimate increased or decreased compared to a standard LMM (Supplementary Figure 4a).

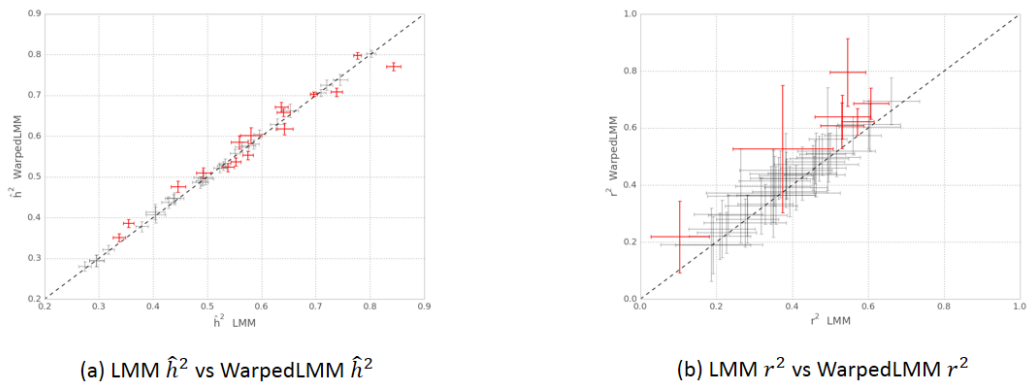

(a) LMM $\hat{h}^2$ vs WarpedLMM $\hat{h}^2$  (b) LMM $r^2$ vs WarpedLMM $r^2$

*Supplementary Figure 4* Comparative analysis of WarpedLMM and a LMM on the yeast dataset. Panel (**a**) shows heritability estimates using a LMM on the untransformed phenotype versus the heritability estimates obtained by WarpedLMM. Empirical error bars were obtained from 10 bootstrap replicates, using 90 % of the data in each replicate. Significant differences are colored in red (paired t-test, α = 0.05). Panel (**b**) shows out-of-sample prediction accuracy assessed by the squared correlation coefficient $r^2$, considering either a LMM on the untransformed data and a WarpedLMM. Prediction accuracies were assessed from 10 random train-test splits. Phenotypes with significant deviations in prediction accuracy of the LMM and the WarpedLMM are highlighted in red (paired t-test, p-value ≤ 0.05).

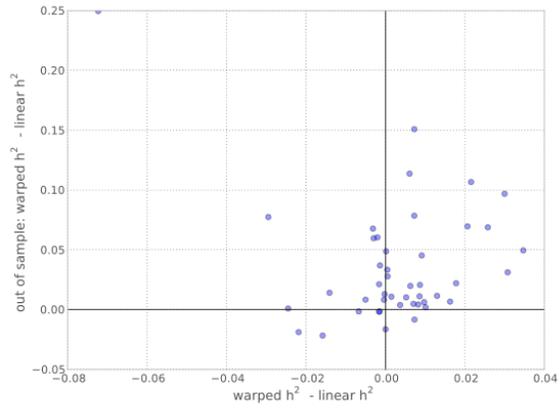
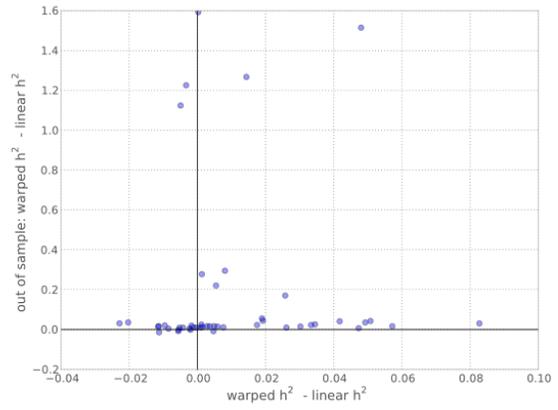

(a) Difference in $\hat{h}^2$ vs difference in $r^2$ in the yeast dataset

(b) Difference in $\hat{h}^2$ vs difference in $r^2$ in the mouse dataset

*Supplementary Figure 5* Comparison of the difference in heritability estimation and the out-of-sample prediction performance in (**a**) the yeast dataset (**b**) the mouse dataset.

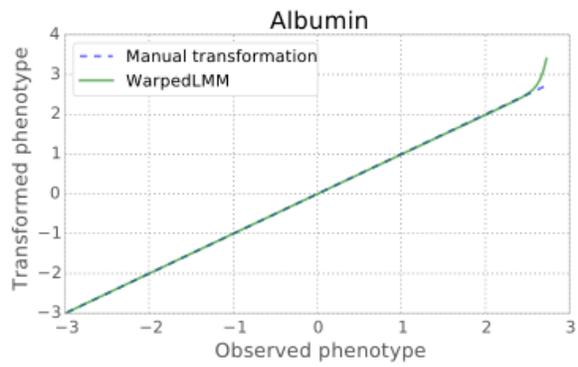 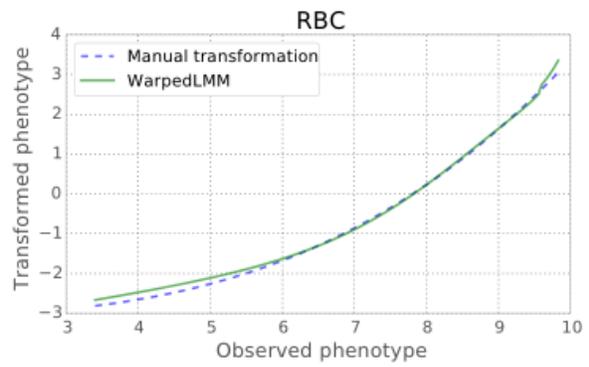 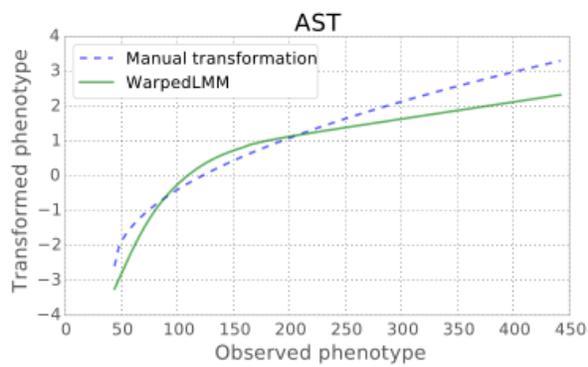 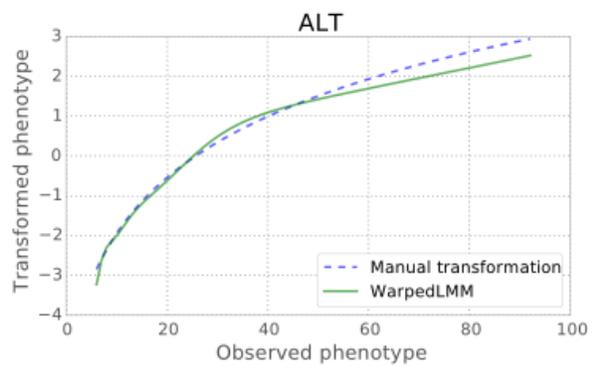

*Supplementary Figure 6* Comparison of the manual transformations reported in (Valdar et al., 2006) and the transformations found by WarpedLMM on the mouse dataset

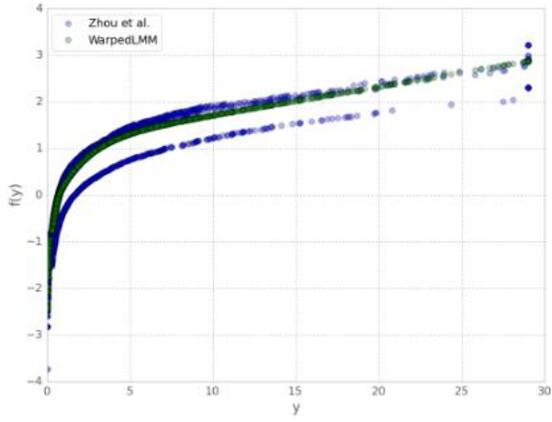
(a) CRP

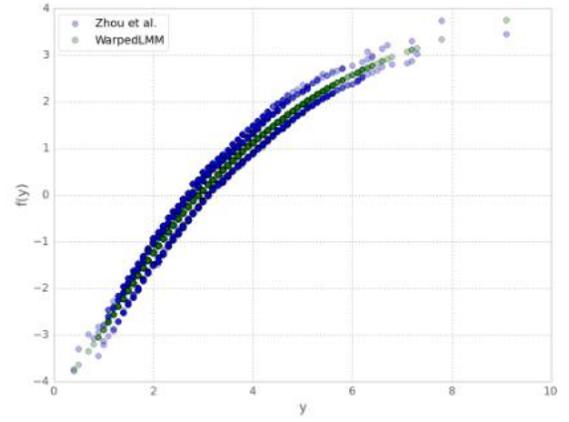
(b) LDL

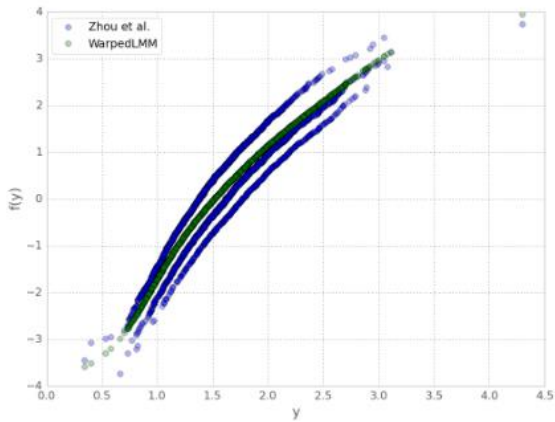
(c) HDL

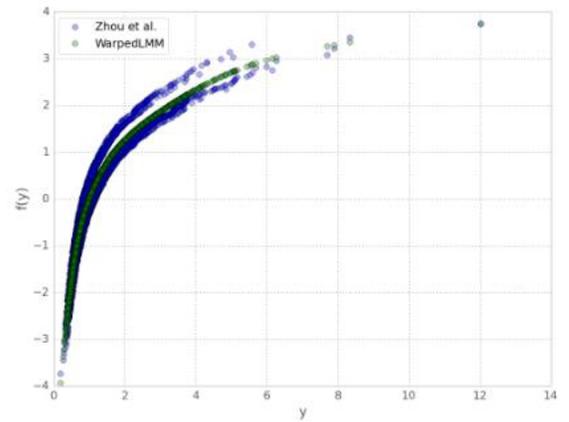
(d) TRI

*Supplementary Figure 7* Comparison of the manual transformations reported in (Zhou & Stephens, 2013) and the transformations found by WarpedLMM on the human dataset

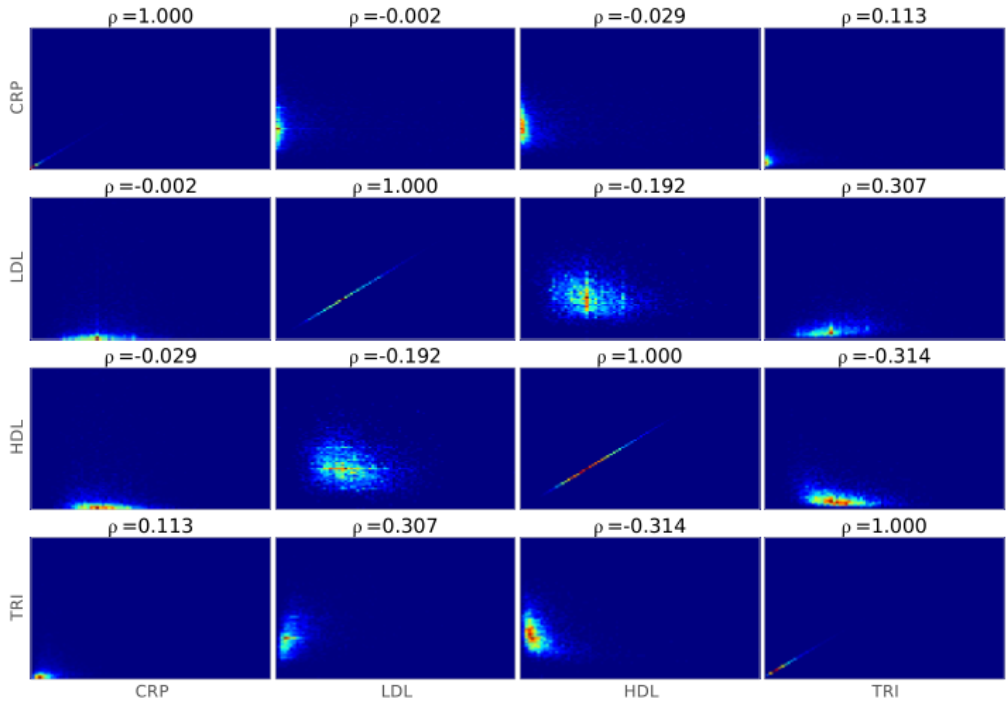

(a) Without transforming the phenotypes

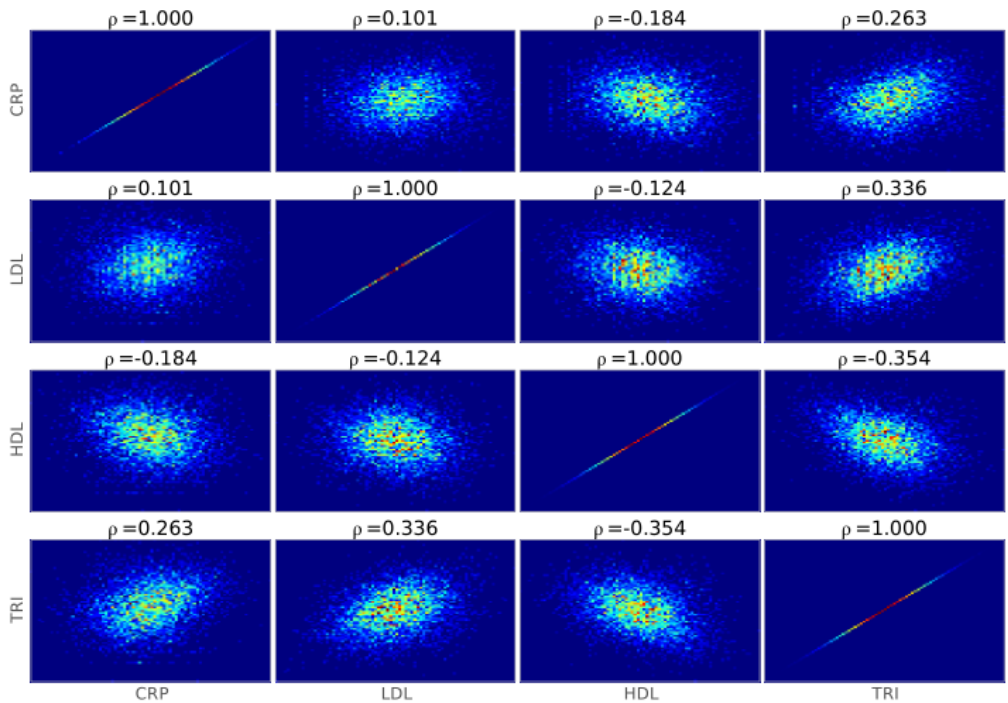

(b) Applying the transformation found by warpedLMM

*Supplementary Figure 8* Correlation between the 4 phenotypes considered in (Zhou & Stephens, 2013) (a) without transforming the phenotypes (b) after applying the transformation reconstructed by WarpedLMM.

*Supplementary Table 1* Association results for the human dataset. Significantly associated loci (at significance level $5 \times 10^{-8}$) have a green background, while non-significant ones are colored in red.

|  | Chr | Position | WarpedLMM | LMM on untransformed | LMM using transformation from original paper (Kang et al., 2010; Sabatti et al., 2009) |
|---|---|---|---|---|---|
| **CRP** | 1 | (157908973, 157966663) | 1.24e-22 | 1.81e-08 | 2.74e-22 |
|  | 12 | (11987334, 119923227) | 1.04e-13 | 1.46e-08 | 3.34e-12 |
| **LDL** | 1 | 55579053 | 3.63e-08 | 1.81e-07 | 1.81e-07 |
|  | 1 | 109620053 | 2.44e-15 | 7.34e-16 | 7.34e-16 |
|  | 1 | 205941798 | 4.21e-08 | 1.74e-07 | 1.74e-07 |
|  | 2 | (21085700, 21165196) | 4.41e-10 | 8.05e-10 | 8.05e-10 |
|  | 19 | 11056030 | 1.99e-08 | 1.49e-08 | 1.49e-08 |
|  | 19 | 50087106 | 6.14e-9 | 1.81e-07 | 1.81e-07 |
| **HDL** | 15 | (56470658, 56478046) | 9.62e-13 | 2.78e-12 | 2.78e-12 |
|  | 16 | (55542640, 55564091) | 4.96e-36 | 1.44e-34 | 1.44e-34 |
|  | 16 | (66229305, 66582496) | 8.11e-09 | 9.79e-09 | 9.79e-09 |
|  | 20 | 42475778 | 3.80e-08 | 2.49e-07 | 2.49e-07 |
| **TRY** | 2 | (27584444, 27594741) | 2.66e-10 | 3.15e-09 | 2.66e-10 |
|  | 8 | 19875201 | 5.57e-09 | 4.08e-08 | 5.57e-09 |